\documentclass[conference, twocolumn]{IEEEtran}                                                  % paper
\IEEEoverridecommandlockouts                              % This command is only
\title{Lossy Source Transmission over the Relay Channel}
\author{
  \authorblockN{Deniz G\"{u}nd\"{u}z \authorrefmark{1}\authorrefmark{2},
     Elza Erkip  \authorrefmark{3}\authorrefmark{1},
     Andrea J. Goldsmith \authorrefmark{2},
     H. Vincent Poor  \authorrefmark{1}
  }\\
    \authorblockA{%
    \authorrefmark{1}Dept.\ of Electrical Engineering,
                     Princeton University, Princeton, NJ, 08544\\
  }
      \authorblockA{%
  \authorrefmark{2}Dept.\ of Electrical Engineering,
                     Stanford University, Stanford, CA 94305
  }
   \authorblockA{%
    \authorrefmark{3}Dept.\ of Electrical and Computer Engineering,
                     Polytechnic University, Brooklyn, NY 11201
  }
  Email:  dgunduz@princeton.edu,elza@poly.edu,andrea@wsl.stanford.edu,poor@princeton.edu
  \thanks{This research was supported in part by the U.S. National Science Foundation under Grants  ANI-03-38807, CCF-04-30885, CCF-06-35177, CCF-07-28208, and CNS-06-25637 and the DARPA's ITMANET program under Grant 1105741-1-TFIND, and the ARO under MURI award W911NF-05-1-0246.}
}
\date{April, 2008}
\usepackage{amsfonts}
\usepackage{amssymb}
\usepackage{amsmath}
\usepackage{amscd}
\usepackage{graphicx}

\newtheorem{thm}{Theorem}[section]

\newtheorem{defn}{Definition}[section]

\begin{document}
%\fontsize{10}{12}\selectfont
\maketitle
\thispagestyle{empty}
\pagestyle{empty}

%%%%%%%%%%%%%%%%%%%%%%%%%%%%%%%%%%%%%%%%%%%%%%%%%%%%%%%%%%%%%%%%%%%%%%%%%%%%%%%%
\begin{abstract}
Lossy transmission over a relay channel in which the relay has access to correlated side information is considered. First, a joint source-channel decode-and-forward scheme is proposed for general discrete memoryless sources and channels. Then the Gaussian relay channel where the source and the side information are jointly Gaussian is analyzed. For this Gaussian model, several new source-channel cooperation schemes are introduced and analyzed in terms of the squared-error distortion at the destination. A comparison of the proposed upper bounds with the cut-set lower bound is given, and it is seen that joint source-channel cooperation improves the reconstruction quality significantly. Moreover, the performance of the joint code is close to the lower bound on distortion for a wide range of source and channel parameters.
\end{abstract}

\section{Introduction}

In many sensor network applications, the goal is to obtain a high fidelity reconstruction of an underlying physical phenomenon at the access point. While sensors might each try to send their own observations directly to the access point, this might result in poor reconstruction quality due to the power limitations of the sensor nodes. Cooperative transmission, in which the nearby nodes help each other's transmissions, has been analyzed extensively in the literature as a promising technique to overcome the power limitation \cite{Cover_ElGamal, Kramer_Gastpar}. However, in many sensor network applications, the sensor observations are highly correlated. It is natural to exploit correlated source coding together with cooperative communication, and this might lead to better reconstruction at the access point than using these two techniques independently. Finding
the optimal strategy in terms of the end-to-end reconstruction fidelity in a general network is a very difficult problem. In fact, we do not even know the necessary and sufficient conditions for lossless transmission over simple network components such as multiple access or broadcast channels with correlated sources \cite{Cover_Salehi}, \cite{Han_Costa}. There are some special cases for which the optimal performance has been characterized \cite{Han}-\cite{Gunduz_IF_MAC}. However, it is likely that the optimal performance requires joint source-channel coding (e.g. \cite{Gunduz_IF_MAC, Gastpar_ITA}), as Shannon's source-channel separation theorem does not apply to multi-terminal scenarios in general.

In this paper, we consider cooperative transmission in which the relay has correlated side information. The goal is to reconstruct the observation of the source terminal at the destination with the least possible distortion. In this scenario, the relay terminal can help the source by both improving the channel transmission rates and by reducing the source compression distortion. In general, the problem is a joint source-channel coding problem, and characterization of the necessary and sufficient conditions is an open problem. We considered this problem for lossless transmission in \cite{ITW07} and gave the necessary and sufficient conditions for some special cases. In \cite{ISIT07}, we focused on the Gaussian relay channel with quadratic Gaussian sources, and provided achievable schemes and comparisons with the joint source-channel cut-set bound. The proposed joint source-channel relaying schemes in \cite{ISIT07} are grouped into three types. In the first group, called \emph{channel cooperation}, the relay simply ignores its side information, and applies one of the well-known cooperative transmission schemes such as decode-and-forward (DF) or compress-and-forward (CF) \cite{Cover_ElGamal}. In the second group, called \emph{source cooperation}, the relay ignores its received signal and uses only its side information either by uncoded transmission followed by MMSE estimation at the decoder, or by a separation-based scheme in which the source and the relay first compress their sources at rates specified by one-helper source coding problem \cite{Oohama} and then transmit the compressed information separately by multiple access channel coding \cite{ISIT07}. Finally, in the third group, called \emph{hybrid cooperation}, the relay sends a superposition of channel and source cooperation codewords by a suitable power allocation. It is shown in \cite{ISIT07} that hybrid cooperation performs better than either of the other two types, and approaches the cut-set lower bound closely for various source-channel settings.

In this paper, we extend the results of \cite{ISIT07} by using the techniques of \cite{ITW07}, which were originally developed for lossless transmission. Specifically, we propose a joint source-channel decode-and-forward (jDF) scheme for discrete memoryless sources and channels. In jDF, the relay uses its side information to improve the relay transmission rates through a joint source-channel coding technique. The results are then generalized to the Gaussian case. In addition to jDF, for the Gaussian case, we propose a joint source-channel partial DF protocol as well as a hybrid version of jDF, in which the relay cooperates for both source compression and channel transmission. Then we propose a generalized source-channel cooperation protocol combining all of the above techniques, which reduces to the previous protocols as special cases. We compare the squared-error distortion achieved by the proposed schemes and the cut-set lower bound for various source-channel parameters. Compared to the classical DF scheme, jDF improves the end-to-end distortion significantly, especially when the relay side information quality is high. Further improvement is obtained with the generalized protocol which dominates the other known schemes for a wide range of source-channel parameters.

\section{System Model}\label{s:sys_model}

In this section, we introduce the system model for discrete memoryless (d.m.) sources and d.m. channels with a general distortion measure and input cost constraint. Later in Section \ref{s:Gaussian}, we concentrate on the Gaussian source and channel case and provide a comparison of the proposed schemes for this case. Let $\{S_{1,k}, S_{2,k}, S_{3,k}\}_{k=1}^\infty$ be a sequence of independent drawings of the correlated random variables $S_1$, $S_2$ and $S_3$ having joint distribution $P_{S_1S_2S_3}$ on the set $\mathcal{S}_1 \times \mathcal{S}_2 \times \mathcal{S}_3$.

The source $S_1$ is to be transmitted over a discrete memoryless relay channel specified by the conditional distribution $P_{YY_1|X_1X_2}$ over the set $\mathcal{X}_1 \times \mathcal{X}_2 \times \mathcal{Y}_1 \times \mathcal{Y}$, where $\mathcal{X}_1, \mathcal{X}_2$ are the input alphabets for the source and the relay, respectively, and $\mathcal{Y}_1, \mathcal{Y}$ are the relay and the destination output alphabets, respectively. The relay terminal has access to the correlated side information $S_2$, and the destination has access to the side information $S_3$. The goal is to reproduce $S_1$ at the destination according to the single letter distortion measure $d: \mathcal{S}_1 \times \mathcal{\hat{S}}_1 \rightarrow [0,\infty)$.

The source encoder observes $S_1^m=(S_{1,1},\ldots,S_{1,m})$ and
maps it to the transmitted codeword $X_1^n=(X_{1,1},\ldots,X_{1,n})$ using a joint source-channel
encoding function $f_1^{(m,n)}$. At time instant $k$, the
received signals at the relay and the destination are $Y_{1,k}$ and $Y_{k}$, respectively. The relay encoder is $f_2^{(m,n)}= \left(f_{2,1}^{(m,n)}, \ldots, f_{2,n}^{(m,n)}\right)$ where
we have $X_{2,k} = f_{2,k}^{(m,n)}(Y_{1,1},\ldots, Y_{1,k-1}, S_2^m)$, for
$1\leq k\leq n$. We denote by $b\triangleq n/m$ the rate of the code.

We assume separate cost constraints $\Gamma_1$, $\Gamma_2$ at the source and the relay, respectively. The average cost function is given as $\gamma_i (X_i^n) = \frac{1}{n} \sum_{j=1}^n \gamma_i(X_{i,j})$, where $\gamma_i : \mathcal{X}_i \rightarrow [0,\infty)$ is the cost of the input signal for $i=1,2$. The destination decoder observes the received vector $Y^n=(Y_1, \ldots, Y_n)$ and its own correlated side information $S_3^m$ and outputs its estimate of the source $\hat{S}_1^m = (\hat{S}_{1,1}, \ldots, \hat{S}_{1,m}) = g^{(m,n)}(Y^n, S_3^m)$, where $g^{(m,n)} : \mathcal{Y}^n \times \mathcal{S}_3^m \rightarrow \mathcal{\hat{S}}_1^m$ is the reconstruction function.

\begin{defn}
For given $(b, \Gamma_1, \Gamma_2)$, we say that the average distortion $D$ is achievable if, for any $\epsilon>0$, there exist sufficiently large $m,n$ with $n/m=b$, and encoding and decoding functions $(f_1^{(m,n)}, f_2^{(m,n)}, g^{(m,n)})$ satisfying the input cost constraints $E \left[ \gamma_i (f_i^{(m,n)}) \right]\leq \Gamma_i$, for $i=1,2$, while the average distortion is bounded as
\begin{equation}
\frac{1}{m} E \left[ \sum_{i=1}^m d(S_{1,i}, \hat{S}_{1,i} )
\right] \leq D +\epsilon.
\end{equation}
\end{defn}

The minimum achievable distortion $D_{min}$ for given $(b, \Gamma_1, \Gamma_2)$ is defined as $D_{min} \triangleq \inf \{
D:D \mbox{ is achievable} \}$. We want to find $D_{min}$ for a given source-channel model. In this paper, we provide
upper and lower bounds on $D_{min}$.

\section{Joint Source-Channel Decode-and-Forward (jDF)}\label{s:results}

%We start with the lower bound to the achievable minimum distortion. We use the cut-set arguments
%for source-channel networks given in \cite{Gastpar2}. The following lemma is the cut-set bound specialized to our source-channel relay channel.

%\begin{lem}\label{cutset}
%If distortion $D$ is achievable for given $(b, \Gamma_1, \Gamma_2)$, then we have
%\begin{eqnarray}
%R_{S_1|S_2, S_3}^{WZ}(D) &\leq &  b I(X_1; Y_1, Y |X_2), \mbox{  and} \\
%R_{S_1| S_3}^{WZ}(D) &\leq &  b I(X_1,X_2;Y ),
%\end{eqnarray}
%for any joint input distribution $p(x_1, x_2)$ satisfying the input cost constraints. Here $R_{X|Y}^{WZ}(D)$ is the Wyner-Ziv rate-distortion function for source $X$ and side information $Y$ available at the decoder \cite{Wyner_Ziv}.
%\end{lem}

%In our previous work \cite{ISIT07}, we proposed several upper bounds for the minimum distortion at the destination. Here we propose some new schemes that improve upon the previous schemes for a wide class of source-channel settings. After we introduce these schemes in the subsections below, we provide a comparison of the performances of these new schemes with the schemes in \cite{ISIT07} and the cut-set lower bounds in Section \ref{s:analysis}.

%\subsection{Joint Source-Channel Decode-and-Forward (jDF)}\label{sub1}

This scheme is based on the well-known decode-and-forward relaying scheme; however, we use the side information at the relay to achieve higher rates over the source-relay channel. We use a joint source-channel relaying technique \cite{ITW07} which was introduced for lossless transmission of a source over the relay channel with correlated relay and destination side information. In the lossy case, the source terminal encoder generates a quantization of the source $S_1$ based on an auxiliary random variable $Z$ satisfying the Markov chain relationship $Z-S_1-(S_2, S_3)$, and transmits a codeword to the destination with the help of the relay. For transmission, we use block Markov encoding. At the destination backward decoding specialized to this  joint source-channel coding is used. We have the following achievability result for the lossy scenario, which we give without proof due to lack of space. We refer the reader to \cite{ITW07} for the details of the scheme.

\begin{thm}\label{t:jDF}
Given a relay channel $p(y, y_1|x_1, x_2)$ and source and side information with joint distribution $p_{S_1 S_2 S_3}$, then distortion $D$ with respect to some distortion measure $d(\cdot, \cdot)$  is achievable for a given triplet $(b,\Gamma_1, \Gamma_2)$ if there exists an auxiliary random variable $Z$ and input distributions $P_{X_1}$ and $P_{X_2}$ at the source and the relay, respectively, satisfying the following conditions:
\begin{enumerate}
  \item $Z-S_1-(S_2, S_3)$ form a Markov chain,
  \item $E[d(S_1, h(Z,S_3))] \leq D$ for some reconstruction function $h: \mathcal{Z}^m \times \mathcal{S}_3^m \rightarrow \mathcal{\hat{S}}_1^m$, and
  \item $X_i$ satisfy the input cost constraints $E[\gamma(X_i)] \leq \Gamma_i$,
  \item \begin{align}
    I(S_1;Z|S_2) \leq& b I(X_1;Y_1|X_2),  \mbox{  and} \\
    I(S_1;Z|S_3) \leq& b I(X_1, X_2; Y),
    \end{align}
\end{enumerate}
where $p(s_1, s_2, s_3, z, x_1, x_2, y_1, y)=p(s_1, s_2, s_3)p(z|s_1)$ $ p(x_1, x_2)p(y_1, y|x_1, x_2)$, and the cardinality bound $|\mathcal{Z}| \leq |\mathcal{S}_1|+2$ is satisfied.
\end{thm}

%\begin{proof}
%Proof can be found in Appendix \ref{App:DF}.
%\end{proof}

\section{The Quadratic Gaussian Case}\label{s:Gaussian}

\begin{figure}
\centering
\includegraphics[width=3.5in]{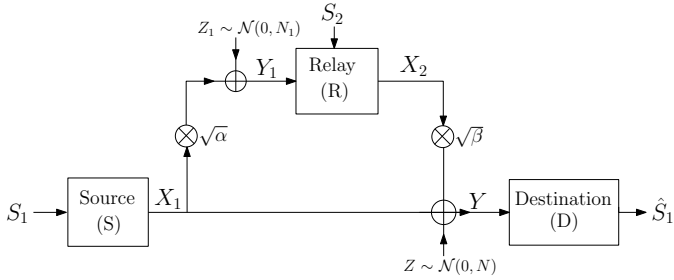}
\caption{Gaussian relay channel with correlated sources
($S_1, S_2$) at the source (S) and the relay (R) terminals.} \label{f:model}
\end{figure}

Here we focus on the quadratic Gaussian problem, in which the source and the side information are jointly Gaussian while the channel is an additive white Gaussian noise (AWGN) channel. For clarity of presentation, we assume that the destination does not have side information, i.e., $\mathcal{S}_3 = \varnothing$, although our achievability schemes can be extended to the case with side information as well. Without loss of generality, we assume that the sequences $\{S_{1,i}\}$ and $\{S_{2,i}\}$ are independent and identically distributed (i.i.d.), zero mean and jointly Gaussian with covariance
\begin{equation}
C_{S_{1}S_{2}} = \left[ \begin{array}{cc}
        1 & \rho\\
        \rho  &  1\\
        \end{array} \right],
\end{equation}
where $\rho \in [-1,1]$ is the correlation coefficient. The distortion is measured by the squared error distortion measure $d(s, \hat{s}) = (s-\hat{s})^2$. The channel is represented as follows:
\begin{eqnarray}
Y_{1,k} &=& h_2 X_{1,k} + Z_{1,k}, \\
Y_{k} &=& h_1 X_{1,k} + h_3 X_{2,k} + Z_{k},
\end{eqnarray}
for $k=1,\ldots,n$, where $Z_1^n=(Z_{1,1},\ldots,Z_{1,n})$ and $Z^n=(Z_1,\ldots,Z_n)$ are i.i.d. zero-mean Gaussian noise vectors with variances $N_1$ and $N$, respectively, and are independent of $X_1^n$, $X_2^n$ and each other. The input cost is measured by the power of the channel input $\gamma_i(x)=x^2$, and we set $\Gamma_i =P_i$, $i=1,2$. Without loss of generality, we assume $|h_1|^2=1$, and define $|h_2|^2 \triangleq \alpha$ and $|h_3|^2 \triangleq \beta$ and let $N=N_1=1$ (see Fig. \ref{f:model}). We assume one channel use per source sample, i.e., $b=1$.

The joint source-channel cut-set lower bound \cite{Gastpar2} in the quadratic Gaussian case is given as \cite{ISIT07}:
\begin{eqnarray}
D_{min} \geq \min_{0 \leq \xi \leq 1} \max \{(1-\rho^2) (1 + (1-\xi^2)(1+\alpha)P_1)^{-1},\nonumber \\
\sigma_1^2 (1 + P_1 + \beta P_2 + 2\xi \sqrt{\beta P_1P_2})^{-1} \}, \nonumber
\end{eqnarray}
in which $\xi$ is the correlation coefficient between the source and the relay codewords.

%---------------------------
%---------------------------

\subsection{Joint Source-Channel Decode-and-Forward (jDF)} \label{ss:jDF}

Using Theorem \ref{t:jDF}, a quantization of the source $S_1$ is transmitted to the destination using an auxiliary source codebook $Z$. Let $Z = S_1 + W$, where $S_1 \perp W$ and $W \sim \mathcal{N}(0, \sigma_W^2)$. Then we have
\begin{align}
    I(S_1;Z|S_2) =& \frac{1}{2} \log \left(1+\frac{1-\rho^2}{\sigma_W^2}\right), \mbox{ and} \\
    I(S_1;Z) =& \frac{1}{2} \log \left(1+\frac{1}{\sigma_W^2}\right).
\end{align}
We also have
\begin{align}
I(X_1;Y_1|X_2)  =& \frac{1}{2} \log (1 + (1-\xi^2)\alpha P_1), \mbox{ and} \nonumber \\
I(X_1, X_2; Y) =& \frac{1}{2} \log (1+ P_1 + \beta P_2 + 2\xi \sqrt{\beta P_1 P_2}),
\end{align}
where, as before, $\xi$ is the correlation coefficient between the channel inputs $X_1$ and $X_2$. For given $P_1, P_2$ and fixed $\xi$, the minimum quantization noise variance is given by
\begin{align}
\sigma_W^2 = & \max \left\{ \frac{1-\rho^2}{(1-\xi^2)\alpha P_1}, \frac{1}{P_1 + \beta P_2 + 2\xi \sqrt{\beta P_1 P_2}} \right\}, \nonumber
\end{align}
and by minimizing over $\xi$ the minimum distortion is found as $D_{jDF}=\min_{0\leq\xi\leq1} \frac{\sigma_W^2}{1+\sigma_W^2}$.

\subsection{Hybrid Joint Source-Channel Decode-and-Forward}\label{ss:hybrid}

In the hybrid joint source-channel DF (hjDF) protocol, the relay divides its power between sending its observation to the destination directly (source cooperation) \cite{ISIT07} and the joint source-channel DF protocol. It then transmits a superposition of the two codewords obtained using the two strategies. Suppose
the relay reserves $\gamma P_2$ ($0\leq \gamma \leq 1$) for the jDF protocol above, and $(1-\gamma) P_2$ for transmitting a quantized version of $S_2$ to the
destination. The destination first decodes the quantized $S_2$, treating the jDF codeword as noise. Then for jDF, this quantized $S_2$ can be used as side information at the destination. Finally, the destination combines the side information received from the relay and the information received from the jDF codeword to obtain a reproduction of the source.

Let $Z_h = S_2 + V$, where $S_2 \perp V$ and $V \sim \mathcal{N}(0, \sigma_V^2)$. The rate at which the relay can send $Z_h$ is
\begin{eqnarray}\label{Rref}
R_h(\gamma) = \frac{1}{2} \log \left(1 + \frac{\beta (1-\gamma) P_2}{1 + P_1 + \beta \gamma P_2}\right).
\end{eqnarray}
Then we have
\begin{align}
\sigma_V^2  \geq& \frac{1 + P_1 + \beta \gamma P_2}{\beta (1-\gamma) P_2}.
\end{align}
For jDF, we set $Z=S_1+W_1$ with $S_1 \perp W_1$ and $W_1 \sim \mathcal{N}(0, \sigma_{W_1}^2)$ in Theorem \ref{t:jDF}. We need to have $I(S_1;Z|S_2) \leq I(X_1; Y_1 | X_2)$  and $I(S_1;Z|Z_h) \leq I(X_1, \bar{X}_2; Y)$, where $X_2$ is the codeword of the relay forwarding $Z_h$. We have
\begin{align}
I(S_1;Z|Z_h) &= \frac{1}{2} \log \left( 1+\frac{1-\rho^2+\sigma_V^2}{\sigma_{W_1}^2(1+\sigma_V^2)} \right).
\end{align}
Other mutual information expressions can be calculated as in Section \ref{ss:jDF} by replacing $P_2$ with $\gamma P_2$. For fixed $\xi$, the minimum quantization noise is found as
\begin{align}
\sigma_{W_1}^2 = \max &  \left\{ \frac{1-\rho^2}{(1-\xi^2)\alpha P_1}, \nonumber \right. \\
 & \left. \frac{1-\rho^2+\sigma_V^2}{(1+\sigma_V^2)(P_1 + \beta \gamma P_2 + 2\xi \sqrt{\beta P_1 \gamma P_2})} \right\}, \nonumber
\end{align}
and the minimum distortion for hjDFh is found as
\[ D_{jDFh} = \min_{0\leq \gamma \leq 1} \min_{0\leq\xi\leq1} \left( \frac{1+\sigma_V^2-\rho^2}{1+\sigma_V^2} + \frac{1}{\sigma_{W_1}^2} \right)^{-1}.\]

%---------------------------
\begin{figure}
\centering
\includegraphics[width=3.5in]{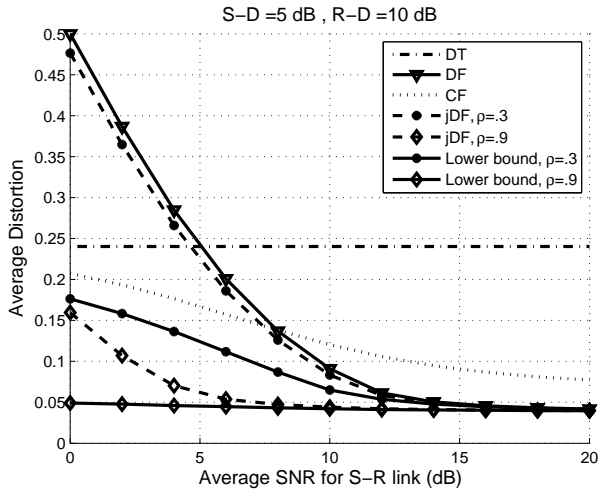}
\caption{Comparison of pure channel cooperation protocols DF and CF, which ignore the side information at the relay, with jDF for low and high quality side information.} \label{f:comp_jDF}
\end{figure}
%---------------------------

\subsection{Joint Source-Channel Partial DF}

In the partial decode-and-forward protocol \cite{Cover_ElGamal}, only a portion of the source message is transmitted using the relay, while the rest is transmitted directly from the source. It is shown in \cite{Zahedi} that, for AWGN channels the partial DF protocol picks the best protocol between direct transmission (DT) and decode-and-forward (DF), but does not improve the achievable rate of DF. However, as we will show, in the presence of correlated side information at the relay terminal, partial joint source-channel DF (pjDF) has the potential for achieving lower distortion.

In pjDF, the source terminal divides its power into two portions. It uses $\theta P_1$ of its power for jDF as in Section \ref{ss:hybrid}, while the rest is used for direct transmission. The two codewords are superimposed. As in partial DF \cite{Cover_ElGamal}, first the jDF portion is decoded by treating the direct transmission as noise. For jDF, the source transmits $Z_2 = S_1+W_1+W_2$ to the destination, where $W_1$ and $W_2$ are independent Gaussian random variables with $S_1 \perp (W_1, W_2)$, $W_1 \sim \mathcal{N}(0, \sigma_{W_1}^2)$ and $W_2 \sim \mathcal{N}(0, \sigma_{W_2}^2)$. It can be shown that, for fixed $\xi$, the minimum quantization noise variance for $Z_2$ is
\begin{align}
\sigma_{W_1}^2 + \sigma_{W_2}^2 \geq  \max  & \left\{ \frac{(1-\rho^2)(1+\alpha(1-\theta)P_1)}{(1-\xi^2)\alpha \theta P_1}, \nonumber \right. \\
 & \left. \frac{1+(1-\theta)P_1}{P_1 + \beta P_2 + 2\xi \sqrt{\beta P_1 P_2}} \right\}.
\end{align}
For direct transmission, the source transmits $S_1+W_1$ utilizing the side information at the destination. Note that the relayed signal is subtracted before decoding the direct information. We need $I(S_1;S_1+W_1 |S_1+W_1+W_2) \leq 1/2\log(1+ (1-\theta) P_1))$. We have
\begin{align}
I(S_1;S_1+W_1 |S_1+W_1+W_2) = \frac{1}{2} \log \left( \frac{1+\frac{1}{\sigma_{W_1}^2}}{1+\frac{1}{\sigma_{W_1}^2+ \sigma_{W_2}^2}} \right). \nonumber
\end{align}
The minimum distortion is found as $D_{jpDF} = \min_{0\leq \theta, \xi \leq 1} \frac{\sigma_{W_1}^2}{1+\sigma_{W_1}^2}$.

In contrast to the transmission rate of partial DF in AWGN channels, as will be seen in Section \ref{s:analysis}, for certain scenarios pjDF strictly improves the distortion performance relative to both direct transmission and jDF.

\subsection{Hybrid Partial jDF}

We also consider a hybrid version of the above partial jDF scheme where both the source and the relay partition their power, and each terminal uses a portion of its power for direct transmission of its source samples. The remaining power at the source and the relay is used for cooperation through the jDF protocol. We call this scheme hybrid partial jDF (hpjDF). The minimum distortion expression for this scheme can be derived as in the previous subsections.

%---------------------------
\begin{figure}
\centering
\includegraphics[width=3.5in]{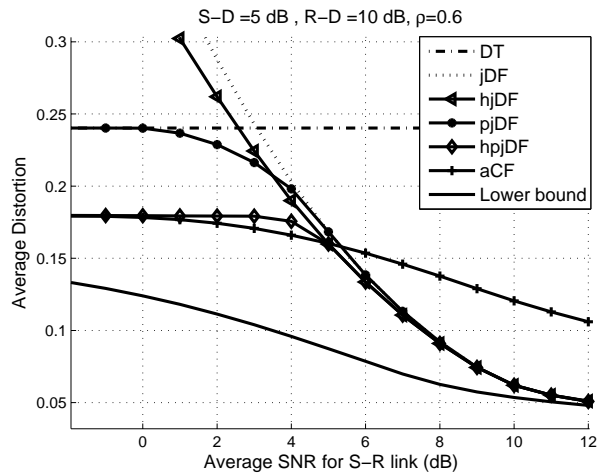}
\caption{Average distortion vs. S-R link quality.} \label{f:SR_jDF}
\end{figure}
%---------------------------

\section{Comparison and Discussion of the results}\label{s:analysis}

In this section, we compare the average distortion
achieved by the strategies in Section \ref{s:Gaussian}. For various source
and channel conditions we also provide comparisons with \cite{ISIT07}. We assume $\sigma_1^2=1$ and measure the $S-R$, $S-D$ and $R-D$ link qualities by $P_1$, $\alpha P_1$ and $\beta P_2$, respectively.

We first analyze the improvement of jDF over the well-known channel cooperation schemes DF and compress-and-forward (CF), where the relay side information is ignored. We set $S-D$ and $R-D$ link qualities to $5$ dB and $10$ dB, respectively. We see in Fig. \ref{f:comp_jDF} that, as expected, jDF improves upon the usual DF even with very low quality relay side information, i.e. for $\rho=0.3$. As the side information quality increases, i.e., for $\rho=0.9$, jDF outperforms both DF and CF, at all channel conditions under consideration. CF performs better than DF when the $S-R$ quality is low.

Next, in Fig. \ref{f:SR_jDF} we compare the different joint source-channel cooperation schemes introduced in Section \ref{s:Gaussian}. We set $S-D$ and $R-D$ link qualities to $5 ~dB$ and $10 ~dB$, respectively and the correlation coefficient to $\rho=0.5$. Note that pjDF improves upon jDF when the S-R link quality is below the $S-D$ link quality. On the other hand hjDF improves upon jDF for a wide range of S-R link SNR's. This improvement increases with the quality of side information, $\rho$, as sending this high quality side information to the destination further helps the reconstruction at the destination. We also include the advanced CF (aCF) protocol, introduced in \cite{ISIT07}, in the plot as well. This is the best CF-based protocol where private information from the source to relay is sent to improve the relay's helper quality while the relay uses both source and channel cooperation through the hybrid strategy. We observe that aCF outperforms other protocols when the $S-R$ quality is relatively low, but jDF-based schemes surpass aCF with increasing $S-R$ quality. hpjDF performs very close to aCF even when the S-R link is weak, and the gap between the upper an lower bounds tightens with increasing $S-R$ quality.

%---------------------------
\begin{figure}
\centering
\includegraphics[width=3.5in]{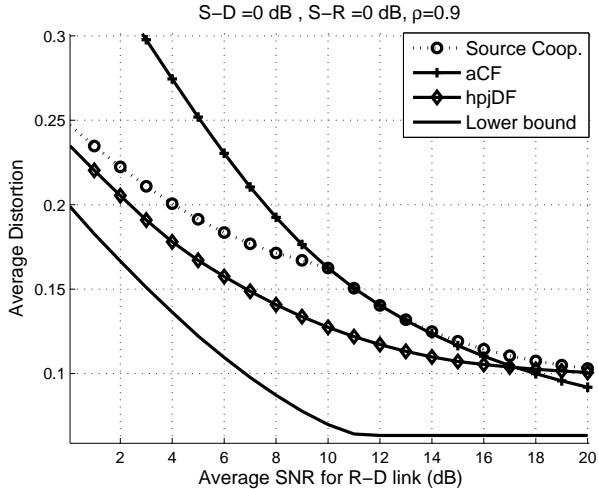}
\caption{Average distortion vs. R-D link quality.} \label{f:RD}
\end{figure}
%---------------------------

Next, we fix both S-D and S-R SNRs at $0$ dB, and consider high
quality side information ($\rho=0.9$). Along with hpjDF we also plot aCF and source cooperation, in which we ignore the received signal at the relay and choose the best among uncoded transmission and a separation based scheme detailed in \cite{ISIT07}. In Fig. \ref{f:RD}, we see that source cooperation performs much better than aCF for low R-D qualities. Due to high correlation and a relatively weak S-R
link, it is more important to transmit relay's side information to the destination. hpjDF outperforms all other protocols up to large R-D SNR values, after which aCF starts to dominate. This is in accordance with the fact that the CF strategy performs better than DF in terms of achievable rates as the S-R link becomes the bottleneck for DF \cite{Kramer_Gastpar}.

Finally, we fix the average received SNRs for S-D, S-R and R-D links
at $4$, $10$ and $4$ dB, respectively, and compare achievable
distortions for increasing side information quality at the relay. For low values of $\rho$ values, aCF performs better than the other protocols. This is due to the relatively high quality of the R-D link which favors CF transmission as seen in Fig. \ref{f:RD} as well. However, with increasing side information quality, both pure source cooperation and hpjDF outperform aCF. Note that, with even $\rho=1$, i.e., when the relay has the exact source information, aCF falls short of the lower bound as it does not utilize its side information for beamforming to the destination as the DF based schemes.

%---------------------------
\begin{figure}
\centering
\includegraphics[width=3.5in]{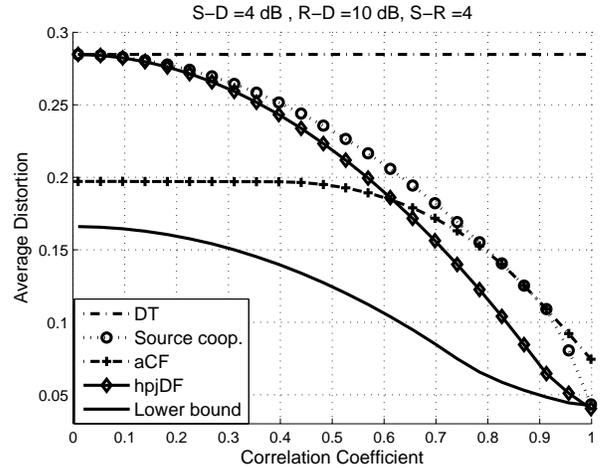}
\caption{Average distortion vs. correlation coefficient ($\rho$) for
fixed links.} \label{cor_vary}
\end{figure}
%---------------------------

\section{Conclusion}\label{Conc}

We have considered joint source-channel coding over the relay channel in the presence of correlated side information at the relay and the destination
terminals. We have proposed a joint source-channel decode-and-forward (jDF) protocol which improves upon the usual decode-and-forward scheme by
using the correlated side information at the relay. Then we have considered the Gaussian relay channel with jointly Gaussian source and relay side
information and analyzed the achievable squared-error distortion at the destination. We have proposed a hybrid partial jDF protocol where the relay allocates its power between sending helper information and jDF while the source allocates
its power between direct transmission and jDF. Currently, we are generalizing our joint source-channel relaying scheme to incorporate compress and forward \cite{Cover_ElGamal} type forwarding at the relay as well.

%We consider transmitting a Gaussian source over a Gaussian relay
%channel where the relay has correlated side information. We propose
%several source-channel coding schemes, and compare achievable
%distortion performances with the lower bound based on cut-set
%arguments. We propose three basic types of strategies: channel
%cooperation, source cooperation and hybrid schemes. In channel
%cooperation, the relay is used only for channel coding and its side
%information is ignored; in source cooperation, the relay is only
%used for its side information ignoring its received signal; and
%hybrid strategies combine these two. The strategy that achieves the
%best performance depends on the correlation coefficient and average
%channel qualities. In particular, we observe that source cooperation
%performs well when correlation is high and the source relay link has
%low quality, while channel cooperation performs well for low
%correlation cases. Hybrid schemes extend the benefits to a wide
%range of correlation and channel conditions, and for most cases
%perform very close to the lower bound.

\end{document}